\documentclass[submitting]{nst}
\makeatletter
\def\NST@lineno{}

\def\linenumbers{}
\makeatother

\usepackage{subfigure,dcolumn}
\usepackage[T2A,T1]{fontenc}
\usepackage[russian,english]{babel}

\usepackage{listings}
\begin{document}

\title{ Rotating and non-linear magnetic-charged black hole with an anisotropic matter field}\thanks{Supported by Yunnan Xingdian Talent Support Program - Young Talent Project, the National Natural Science Foundation of China (Grant No. 12463001)}

\author{Qi-Quan Li}
\address{School of Physics Science and Technology, Xinjiang University, Urumqi 830046, China}

\author{Yu Zhang} 
\email[Corresponding author:~]{zhangyu\_128@126.com}
\address{Faculty of Science, Kunming University of Science and Technology, Kunming 650500, China}

\author{Hoernisa Iminniyaz} 
\email[Corresponding author:~]{wrns@xju.edu.cn}
\address{School of Physics Science and Technology, Xinjiang University, Urumqi 830046, China}

	\begin{abstract}
		We present the solution of a non-linear magnetic-charged black hole with an anisotropic matter field and further extend it to obtain the corresponding rotating black hole solution using the modified Newman-Janis algorithm. The event horizon and ergosphere of the rotating black hole are studied in terms of the perspective of geometric properties, revealing that the rotating black hole can have up to three horizons. The first law of thermodynamics and the squared-mass formula for the rotating black hole are derived from a thermodynamic perspective, based on which we obtain the thermodynamic quantities and study the thermodynamic stability of the rotating black hole. Additionally, we calculate the Penrose process for the rotating black hole, indicating the influence of various black hole parameters on the maximal efficiency of the Penrose process.
		
	\end{abstract}


	\maketitle

\section{Introduction}
	From the prediction of black holes by general relativity to the release of the first image of a black hole shadow in Virgo A* galaxy (M87) by the Event Horizon Telescope Collaboration in April 2019 \cite{EventHorizonTelescope:2019dse}, more than a hundred years have passed. Although we have made significant advancements in both observation and black hole theory, there still remains the unavoidable issue of singularities in black hole theory. The singularity of a black hole possesses a series of peculiar properties, with infinite density and curvature, where the currently known physical laws fail \cite{Hawking:1973uf}. The black hole singularity problem originates from general relativity, but general relativity is not applicable at the singularity, revealing the incomplete development of our current gravity theory. It is generally believed that the solution to this problem can only be achieved through a complete theory of quantum gravity, where the black hole singularity can be fully resolved. However, we have not yet discovered a complete theory of quantum gravity, and therefore using semiclassical theories to address the singularity problem becomes extremely valuable.

In 1968, Bardeen presented the first solution for a regular black hole \cite{bardeen1968non}, which avoids the singularity by incorporating a de Sitter core that generates negative pressure, thereby preventing a singular end-state of the gravitationally collapsed matter \cite{Sakharov:1966aja,Nam:2018uvc}. Afterwards, Ay\'{o}n-Beato and Garc\'{i}a considered the coupling between the gravitational field and a non-linear electromagnetic field \cite{Ayon-Beato:2000mjt}, and then Bardeen black hole solution could be regained, indicating that the Bardeen model can be explained as a non-linear magnetic monopole in non-linear electrodynamics. Subsequently, more regular black holes were obtained by coupling non-linear electrodynamics with gravitational theory, such as Ay\'{o}n-Beato and Garc\'{i}a black hole \cite{Ayon-Beato:1998hmi}, Hayward black hole \cite{Hayward:2005gi}, and Berej-Matyjasek-Trynieki-Wornowicz black hole \cite{Berej:2006cc}.

In astrophysical research, the spherically symmetric model is one of the simplest models and can be divided into isotropic and anisotropic spherically symmetric celestial models based on whether the pressure is isotropic or anisotropic. Currently, most research focuses on isotropic fluids because astrophysical observations support isotropy \cite{Delgaty:1998uy,Semiz:2008ny}. For example, the perfect Pascalian-fluid (isotropic-fluid) assumption is supported by theory and observation \cite{Vogt:2009gs,Vogt:2010ad,Garcia-Reyes:2018eju}.

However, compared to isotropic models, studies on anisotropic models are relatively limited. Increasing theoretical evidences suggest that various interesting physical phenomena may occur within certain density ranges, leading to local anisotropy \cite{Herrera:1997plx}. This has led to an increase in research on anisotropy. For instance, at the galactic level, dark force effects allow for an effective description in terms of general relativity sourced by an anisotropic fluid \cite{Cadoni:2017evg}. Local anisotropy in self-gravitating systems affects the physical properties, stability, and structure of stellar matter \cite{Bowers:1974tgi,Mak:2001eb,DevK}.

The electromagnetic field outside the Reissner-Nordstr\"{o}m (R-N) black hole is the simplest anisotropic matter field, where the equation of state satisfies $ p_{r}=-\varepsilon$ in the radial direction and $p_{\theta}=p_{\phi}=\varepsilon$ in the angular directions in a local reference frame. Cho and Kim extended the R-N black hole solution to obtain a Schwarzschild black hole with an anisotropic matter field \cite{Cho:2017nhx}. The anisotropic matter field in this black hole is described by parameters $\omega $ and $K$, with the relationship $r_{0}^{2 \omega _{2}} \equiv\left(1-2 \omega _{2}\right) K \geq 0$. Subsequently, a rotating charged black hole with an anisotropic matter field \cite{Kim:2019hfp} was proposed, and its geometry, thermodynamic properties, and energy extraction were studied. For additional research related to an anisotropic matter field, one can refer to Refs.\cite{Jeong:2023hom,Kim:2021vlk,Khodadi:2021mct,Lee:2021sws,AhmedRizwan:2020sza,Badia:2020pnh}.

We have reviewed the research and development work on regular black holes and an anisotropic matter field. In this paper, we consider both to obtain the solution of a non-linear magnetic-charged black hole with an anisotropic matter field and the corresponding rotating black hole. We then further study the geometric properties and thermodynamic properties of this rotating black hole.

The structure of this paper is as follows. In Sec.~\ref{sec:2}, the solution for a non-linear magnetic-charged black hole with an anisotropic matter field is obtained. In Sec.~\ref{sec:3}, we extend the solution using the modified Newman-Janis algorithm to obtain a solution for a rotating and non-linear magnetic-charged black hole with an anisotropic matter field and study its energy-momentum tensor and weak energy condition. In Sec.~\ref{sec:4} and \ref{sec:5} explore the geometric and thermodynamic properties of the rotating black hole, respectively. In Sec.~\ref{sec:6}, we examine the efficiency of the Penrose process for extracting energy from the rotating black hole. Finally, in Sec.~\ref{sec:7}, we present the conclusion and discussion of this study.

\section{Non-linear magnetic-charged black hole with an anisotropic matter field}
\label{sec:2}
We consider the coupling between Einstein gravity and a non-linear electromagnetic field in the presence of an anisotropic matter field, which is described by the following equations
\begin{equation}\label{eq:G1}
	G_{\mu}^{\nu}=2\left[\frac{\partial \mathcal{L}(F)}{\partial F} F_{\mu \rho} F^{v \rho}-\delta_{\mu}^{\nu} \mathcal{L}(F)\right] +8\pi  T_{\mu}^{\nu},
\end{equation}
\begin{equation}\label{eq:G2}
\nabla_{\mu}\left(\frac{\partial \mathcal{L}(F)}{\partial F} F^{v \mu}\right)=0 ,
\end{equation}
\begin{equation}\label{eq:G3}
\nabla_{\mu} * F^{\nu \mu}=0.
\end{equation}
Here, $ \mathcal{L}$ is a function of the invariant $ F \equiv \frac{1}{4} F_{\mu \nu} F^{\mu \nu}$ with $F_{\mu \nu}=\partial_{\mu} A_{\nu}-\partial_{\nu} A_{\mu}$, given by \cite{Nam:2018uvc,Benavides-Gallego:2018odl}
\begin{equation}\label{eq:L}
	\begin{array}{c}
		\mathcal{L}(F)=\frac{3 M}{|Q|^{3}} \frac{\left(2 Q^{2} F\right)^{3 / 2}}{\left[1+\left(2 Q^{2} F\right)^{3 / 4}\right]^{2}},
	\end{array}
\end{equation}
where $Q$ is the magnetic charge. In this work, we consider the black holes with an anisotropic matter field. The energy-momentum tensor of an anisotropic matter field obtained from Ref. \cite{Cho:2017nhx} is given by
\begin{equation}\label{eq:energy}
	\begin{array}{c}
		T_{\nu}^{\mu}=\operatorname{diag}\left(-\varepsilon, p_{r}, p_{\theta}, p_{\phi}\right),
	\end{array}
\end{equation}
where $p_{r}=-\varepsilon(r)$ and $p_{\theta}=p_{\phi}=\omega \varepsilon(r)$. The energy density is
\begin{equation}\label{eq:energy2}
	\begin{array}{c}
		\varepsilon(r)=\frac{r_{0}^{2 \omega }}{8 \pi r^{2 \omega +2}},
	\end{array}
\end{equation}
where $r_0$ is a charge-like quantity of dimension of length and defined by $r_{0}^{2 \omega }=(1-2 \omega ) K$.

To obtain a static spherically symmetric black hole solution, we assume the line element is
\begin{equation}\label{eq:DS1}
d s^{2}=-f(r) d t^{2}+f(r)^{-1} d r^{2}+r^{2} d \Omega^{2},		
\end{equation}
\begin{equation}\label{eq:DS2}
 f(r)=1-\frac{2 m(r)}{r} \text { and } d \Omega^{2}=d \theta^{2}+\sin ^{2} \theta d \phi^{2},
\end{equation}
and use the ansatz for Maxwell field
\begin{equation}\label{eq:F1}
	\begin{aligned}
		\begin{array}{c}
 F_{\mu \nu}=\left(\delta_{\mu}^{\theta} \delta_{\nu}^{\varphi}-\delta_{\nu}^{\theta} \delta_{\mu}^{\varphi}\right) B(r, \theta).
		\end{array}
	\end{aligned}
\end{equation}
From Eqs.(\ref{eq:G2}) and (\ref{eq:G3}), we simplify Eq.(\ref{eq:F1}) to obtain
\begin{equation}\label{eq:F2}
	\begin{aligned}
		\begin{array}{c}
 F_{\mu \nu}=\left(\delta_{\mu}^{\theta} \delta_{\nu}^{\varphi}-\delta_{\nu}^{\theta} \delta_{\mu}^{\varphi}\right) Q \sin \theta,
		\end{array}
	\end{aligned}
\end{equation}
where $Q$ is the integration constant. Further, we obtain $F=Q^{2} / 2 r^{4}$. Note that, the magnetic charge $Q$ is defined in the following integral form
\begin{equation}\label{eq:Q}
	\begin{aligned}
		\begin{array}{c}
 \frac{1}{4 \pi} \int_{S_{2}^{\infty}} \boldsymbol{F}=Q,
		\end{array}
	\end{aligned}
\end{equation}
where $S_{2}^{\infty}$ is a two-sphere at the infinity. From the above results, it is possible to simplify the time component of Eq.(\ref{eq:G1}) as
\begin{equation}\label{eq:TT}
	\begin{aligned}
		\begin{array}{c}
\frac{2}{r^{2}} \frac{\mathrm{d} m(r)}{\mathrm{d} r}=\frac{6 M Q^{3}}{\left(r^{3}+Q^{3}\right)^{2}}+\frac{(1-2\omega )K}{2r^{2\omega+2 }}.
		\end{array}
	\end{aligned}
\end{equation}
By integrating Eq.(\ref{eq:TT}) from $r$ to $\infty$ and using the relation $M=\lim\limits_{r \rightarrow \infty}\left(m(r)-\frac{K}{2} r^{-2\omega +1}\right)$, one finally obtains
\begin{equation}\label{eq:ds2}
	\begin{aligned}
		\begin{array}{c}
			f(r)=1-\frac{2 M r^{2}}{r^{3}+Q^{3}}-\frac{K}{r^{2\omega }}.
		\end{array}
	\end{aligned}
\end{equation}
For the above equation for $K=0$, we can derive the metric function $f(r)$ of the Hayward-like black hole \cite{Hayward:2005gi}. When $Q=0$ and $\omega=1$, we can derive the metric function $f(r)$ of the R-N black hole. It is found that $K$ and $q$ ($q$ is the charge in the R-N black hole) have similar status in the metric function $f(r)$. Therefore, $K$ is the global charge, and $r_0$ is a charge-like quantity of dimension of length with $r_{0}^{2\omega} = (1 - 2\omega)K$.

In the introduction, we mentioned the non-linear magnetic-charged black hole, which is a regular black hole without curvature singularity. Now, we consider whether the anisotropic matter field in the non-linear magnetic-charged black hole with an anisotropic matter field will alter this property. Therefore, we calculate the curvature of this black hole as follows
\begin{equation}\label{eq:R1}
	\begin{aligned}
		\begin{array}{c}
			R=\frac{12 M Q^3 \left(2 Q^3-r^3\right)}{\left(Q^3+r^3\right)^3}+2 K (\omega -1) (2 \omega -1) r^{-2 (\omega +1)},
		\end{array}
	\end{aligned}
\end{equation}
\begin{equation}\label{eq:R2}
	\begin{aligned}
&R_{\mu \nu} R^{\mu \nu}=\frac{72 M^2 Q^6 \left(2 Q^6-2 Q^3 r^3+5 r^6\right)}{\left(Q^3+r^3\right)^6}\\
&+\frac{24 K M Q^3 (2 \omega -1) r^{-2 (\omega +1)} \left(Q^3 (\omega -1)-r^3 (2 \omega +1)\right)}{\left(Q^3+r^3\right)^3}\\
&+2 \left(\omega ^2+1\right) (K-2 K \omega )^2 r^{-4 (\omega +1)},
	\end{aligned}
\end{equation}
\begin{equation}\label{eq:R3}
	\begin{aligned}
		&R_{\mu \nu \sigma \tau} R^{\mu \nu \sigma \tau} = 4K^2 \left(4 \omega ^4+4 \omega ^3+5 \omega ^2+1\right) r^{-4 (\omega +1)} \\
		&\quad + \frac{16 K M }{\left(Q^3+r^3\right)^3 r^{2 (\omega +1)}} \left[Q^6 \left(2 \omega ^2-3 \omega +1\right) \right. \\
		&\quad \left. + Q^3 r^3 \left(-14 \omega ^2-9 \omega +2\right) + r^6 \left(2 \omega ^2+3 \omega +1\right)\right] \\
		&\quad + \frac{48 M^2 \left(2 Q^{12}-2 Q^9 r^3+18 Q^6 r^6-4 Q^3 r^9+r^{12}\right)}{\left(Q^3+r^3\right)^6}.
	\end{aligned}
\end{equation}
From the above results, we find that the curvature diverges at $r=0$ due to the influence of the anisotropic matter field, which indicates that the non-linear magnetic-charged black hole with an anisotropic matter field is not a regular black hole and has a singularity.
	
	\section{Rotating and non-linear magnetic-charged black hole with an anisotropic matter field}
\label{sec:3}

In 1965, the Newman-Janis algorithm (NJA) \cite{Newman:1965tw} was first proposed by Newman and Janis. This algorithm obtains the Kerr metric by performing a complex coordinate transformation on the Schwarzschild line element. The NJA has been widely applied in the scientific community \cite{Kim:2019hfp,Kiselev:2002dx,Liu:2020ola,Toshmatov:2017zpr,Shaikh:2019fpu,Xu:2020jpv,Kumar:2017qws,Xu:2016jod,Benavides-Gallego:2018odl,Tsukamoto:2014tja,Tsukamoto:2017fxq}. Later, Azreg-A\"{i}nou modified the NJA \cite{Azreg-Ainou:2014pra,Azreg-Ainou:2014}, and the only difference from the original one is that they skipped the complexification of coordinates.

In this section, we will use the modified NJA to generalize the solution of the spherically symmetric non-linear magnetic-charged black hole with an anisotropic matter field to obtain the corresponding rotating black hole solution. Let's first review the modified NJA to obtain a general rotating black hole line element from a general static spherically symmetric black hole line element. The line element for a static spherically symmetric spacetime in Boyer-Lindquist (BL) coordinates is
\begin{equation}\label{eq:A1}
	\begin{aligned}
		d s^{2}=-f(r) d t^{2}+g^{-1}(r) d r^{2}+h(r) d \Omega^{2},
	\end{aligned}
\end{equation}
where
\begin{equation}\label{eq:A2}
	\begin{aligned}
d \Omega^{2}=d \theta^{2}+\sin ^{2} \theta d \phi^{2}.
	\end{aligned}
\end{equation}
Considering the following coordinate transformation
\begin{equation}\label{eq:A3}
	\begin{aligned}
d u=d t-\frac{d r}{\sqrt{f(r) g(r)}},
	\end{aligned}
\end{equation}
the line element in Eddington-Finkelstein (EF) coordinates $(u, r, \theta, \phi)$  is obtained as
\begin{equation}\label{eq:A4}
	\begin{aligned}
d s^{2}=-f(r) d u^{2}-2 \sqrt{\frac{f(r)}{g(r)}} d u d r+h(r) d \Omega^{2}.
	\end{aligned}
\end{equation}
In terms of the null tetrads satisfy the relations $l_{\mu} l^{\mu}=n_{\mu} n^{\mu}=m_{\mu} m^{\mu}=l_{\mu} m^{\mu}=   n_{\mu} m^{\mu}=0$ and $l_{\mu} n^{\mu}=-m_{\mu} \bar{m}^{\mu}=1$, the contravariant metric tensor in Eq.(\ref{eq:A4}) can be expressed in terms of the null tetrad as
\begin{equation}\label{eq:A6}
	\begin{aligned}
g^{\mu \nu}=-l^{\mu} n^{\nu}-l^{\nu} n^{\mu}+m^{\mu} \bar{m}^{\nu}+m^{\nu} \bar{m}^{\mu},
	\end{aligned}
\end{equation}
where
\begin{equation}\label{eq:A7}
    \begin{aligned}
        l^{\mu} & = \delta_{r}^{\mu}, \\
        n^{\mu} & = \sqrt{\frac{g(r)}{f(r)}} \delta_{r}^{\mu} - \frac{f(r)}{2} \delta_{t}^{\mu}, \\
        m^{\mu} & = \frac{1}{\sqrt{2 h(r)}} \delta_{\theta}^{\mu} + \frac{i}{\sqrt{2 h(r)} \sin \theta} \delta_{\phi}^{\mu}, \\
        \bar{m}^{\mu} & = \frac{1}{\sqrt{2 h(r)}} \delta_{\theta}^{\mu} - \frac{i}{\sqrt{2 h(r)} \sin \theta} \delta_{\phi}^{\mu}.
    \end{aligned}
\end{equation}
The key step of the modified NJA is the consideration of complex coordinate transformations in the $u-r$ plane
\begin{equation}\label{eq:A8}
    \begin{aligned}
        u \rightarrow u-i a \cos \theta, \hspace{0.3cm}   r \rightarrow r+i a \cos \theta.
    \end{aligned}
\end{equation}
As a result of the transformation of Eq.(\ref{eq:A7}), the metric function also takes a new form: $f(r)$ $\rightarrow$ $F(r, a, \theta)$, $g(r)$ $\rightarrow G(r, a, \theta)$, and $h(r)$ $\rightarrow$ $\Sigma=r^{2}+a^{2} \cos ^{2}\theta$ \cite{Azreg-Ainou:2014pra,Azreg-Ainou:2014}. Thus, the new form of null tetrads is given by
\begin{equation}\label{eq:A9}
    \begin{aligned}
       l^{\mu} & =\delta_{r}^{\mu}, \\
n^{\mu} & =\sqrt{\frac{G}{F}} \delta_{u}^{\mu}-\frac{1}{2} F \delta_{r}^{\mu}, \\
m^{\mu} & =\frac{1}{\sqrt{2 \Sigma}}\left[\delta_{\theta}^{\mu}+i a\left(\delta_{u}^{\mu}-\delta_{r}^{\mu}\right) \sin \theta+\frac{i}{\sin \theta} \delta_{\phi}^{\mu}\right], \\
\bar{m}^{\mu} & =\frac{1}{\sqrt{2 \Sigma}}\left[\delta_{\theta}^{\mu}-i a\left(\delta_{u}^{\mu}-\delta_{r}^{\mu}\right) \sin \theta-\frac{i}{\sin \theta} \delta_{\phi}^{\mu}\right].
    \end{aligned}
\end{equation}
Then, from Eq.(\ref{eq:A6}), the contravariant components of the metric $g^{\mu\nu}$ can be obtained as
\begin{equation}\label{eq:A10}
    \begin{aligned}
       g^{u u} & =\frac{a^{2} \sin ^{2} \theta}{\Sigma}, \\
g^{u r} & =-\sqrt{\frac{G}{F}}-\frac{a^{2} \sin ^{2} \theta}{\Sigma}, \\
g^{u \phi} & =\frac{a}{\Sigma}, \\
g^{r r} & =G+\frac{a^{2} \sin ^{2} \theta}{\Sigma}, \\
g^{r \phi} & =-\frac{a}{\Sigma}, \\
g^{\theta \theta} & =\frac{1}{\Sigma}, \\
g^{\phi \phi} & =\frac{1}{\Sigma \sin ^{2} \theta}.
    \end{aligned}
\end{equation}
The following transformation is used to convert from EF coordinates to BL coordinates:
\begin{equation}\label{eq:A11}
    \begin{aligned}
       d u=d t+\lambda(r) d r, \quad d \varphi=d \phi+\chi(r) d r,
    \end{aligned}
\end{equation}
where the functions $\lambda(r)$ and $\chi(r)$ can be determined by ensuring that all the nondiagonal components of the metric tensor, except the coefficient $g_{t \phi}\left(g_{\phi t}\right)$, are equal to zero. Therefore, we obtain that \cite{Azreg-Ainou:2014pra}
\begin{equation}\label{eq:A12}
    \begin{aligned}
       \lambda(r)=-\frac{k(r)+a^{2}}{g(r) h(r)+a^{2}}, \quad \chi(r)=-\frac{a}{g(r) h(r)+a^{2}},
    \end{aligned}
\end{equation}
with
\begin{equation}\label{eq:A13}
    \begin{aligned}
       k(r)=\sqrt{\frac{g(r)}{f(r)}} h(r),
    \end{aligned}
\end{equation}
and
\begin{equation}\label{eq:A14}
    \begin{aligned}
       F(r, \theta)&=\frac{\left(g h+a^{2} \cos ^{2} \theta\right) \Sigma}{\left(k^{2}+a^{2} \cos ^{2} \theta\right)^{2}}, \\
       \quad G(r, \theta)&=\frac{g h+a^{2} \cos ^{2} \theta}{\Sigma}.
    \end{aligned}
\end{equation}
Finally, the rotating solution corresponding to the spherically symmetric line element Eq.(\ref{eq:A1}) is obtained as follows:
\begin{equation}\label{eq:A15}
    \begin{aligned}
       d s^{2}= & -\frac{\left(g h+a^{2} \cos ^{2} \theta\right) \Sigma}{\left(k+a^{2} \cos ^{2} \theta\right)^{2}} d t^{2}+\frac{\Sigma}{g h+a^{2}} d r^{2} \\
       & -2 a \sin ^{2} \theta\left[\frac{k-g h}{\left(k+a^{2} \cos ^{2} \theta\right)^{2}}\right] \Sigma d \phi d t+\Sigma d \theta^{2} \\
       & +\Sigma \sin ^{2} \theta\left[1+a^{2} \sin ^{2} \theta \frac{2 k-g h+a^{2} \cos ^{2} \theta}{\left(k+a^{2} \cos ^{2} \theta\right)^{2}}\right] d \phi^{2}.
    \end{aligned}
\end{equation}
In the above, we reviewed the modified NJA algorithm which provides a general expression for a rotating black hole metric. Considering the metric function Eq.(\ref{eq:ds2}) given in the previous section, we obtain
\begin{equation}\label{eq:A16}
    \begin{aligned}
       g(r)&=f(r)=1-\frac{2 M r^{2}}{r^{3}+Q^{3}}-\frac{K}{r^{2\omega }}, \\
h(r)&=k(r)=r^{2}.
    \end{aligned}
\end{equation}
Substituting the above equation into Eq.(\ref{eq:A15}), we obtain the line element of a rotating and non-linear magnetic-charged black hole with an anisotropic matter field as
\begin{equation}\label{eq:A17}
    \begin{aligned}
       d s^{2}= & -\left[1-\frac{2 \rho r}{\Sigma}\right] d t^{2}+\frac{\Sigma}{\Delta} d r^{2}-\frac{4 a \rho r \sin ^{2} \theta}{\Sigma} d t d \phi \\
& +\Sigma d \theta^{2}+\sin ^{2} \theta\left[r^{2}+a^{2}+\frac{2 a^{2} \rho r \sin ^{2} \theta}{\Sigma}\right] d \phi^{2},
    \end{aligned}
\end{equation}
where
\begin{equation}\label{eq:A18}
    \begin{aligned}
      \Delta & =r^{2}-2 \rho r+a^{2} \\
\Sigma & =r^{2}+a^{2} \cos ^{2} \theta \\
2 \rho & =\frac{2 M r^{3}}{r^{3}+Q^{3}}+K r^{-2\omega  +1}.
    \end{aligned}
\end{equation}

\begin{figure*}
	\begin{tabular}{c c}
		\includegraphics[scale=0.83]{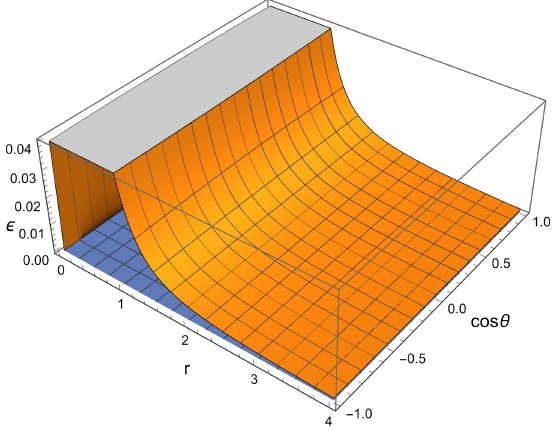}\hspace{-0.2cm}
		&\includegraphics[scale=0.9]{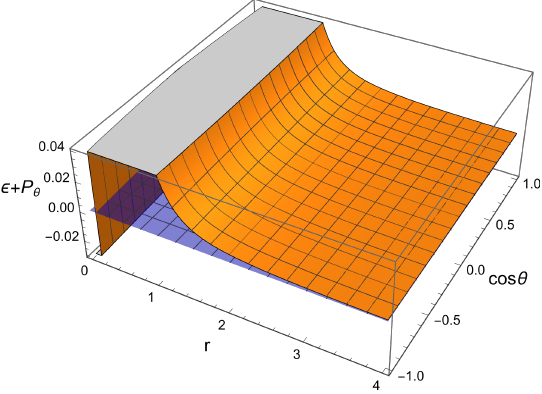}
	\end{tabular}
	\caption{Dependence of matter density $\epsilon$ and $\epsilon+p_{\theta}$ on radius and angle for rotating and nonlinear magnetic-charged black hole with an anisotropic matter field for $M=1$, $Q=0.5$, $a=0.2$, $K=1$, $\omega =0.2$.}\label{fig:energy condition1}
\end{figure*}
\subsection{The energy-momentum tensor}

The nonvanishing components of the Einstein tensor $G_{\mu \nu}$ are given by
\begin{equation}\label{eq:A19}
    \begin{aligned}
      G_{t t} & =\frac{2\left(r^{4}+a^{2} r^{2}-2 r^{3} \rho-a^{4} \sin ^{2} \theta \cos ^{2} \theta\right) \rho^{\prime}}{\Sigma^{3}} \\
& -\frac{a^{2} r \sin ^{2} \theta \rho^{\prime \prime}}{\Sigma^{2}}, \\
G_{t \phi} & =\frac{2 a \sin ^{2} \theta\left[\left(a^{2}+r^{2}\right)\left(a^{2} \cos ^{2} \theta-r^{2}\right)+2 r^{3} \rho\right] \rho^{\prime}}{\Sigma^{3}} \\
& +\frac{a r \sin ^{2} \theta\left(a^{2}+r^{2}\right) \rho^{\prime \prime}}{\Sigma^{2}}, \\
G_{r r} & =-\frac{2 r^{2} \rho^{\prime}}{\Sigma \Delta}, \\
G_{\theta \theta} & =-\frac{2 a^{2} \cos ^{2} \theta \rho^{\prime}}{\Sigma}-r \rho^{\prime \prime}, \\
G_{\phi \phi} & =-\frac{a^{2} \sin ^{2} \theta\left(a^{2}+r^{2}\right)\left(a^{2}+\left(a^{2}+2 r^{2}\right) \cos 2 \theta\right) \rho^{\prime}}{\Sigma^{3}} \\
& -\frac{4 a^{2} r^{3} \sin ^{4} \theta \rho \rho^{\prime}}{\Sigma^{3}}-\frac{r \sin ^{2} \theta\left(a^{2}+r^{2}\right)^{2} \rho^{\prime \prime}}{\Sigma^{2}},
    \end{aligned}
\end{equation}
where the prime $\prime$ denotes the derivative with respect to $r$ and
\begin{equation}\label{eq:A20}
    \begin{aligned}
      2 \rho & =\frac{2 M r^{3}}{r^{3}+Q^{3}}+K r^{-2\omega  +1} ,\\
       2 \rho^{\prime} & =\frac{6 M Q^3 r^2}{\left(Q^3+r^3\right)^2}+K (1-2 \omega ) r^{-2 \omega }, \\
     2 \rho^{\prime \prime}&=\frac{12 M Q^3 r \left(Q^3-2 r^3\right)}{\left(Q^3+r^3\right)^3}-2 K \omega  (1-2\omega ) r^{-2 \omega -1}.
    \end{aligned}
\end{equation}
In order to obtain the components of the energy-momentum tensor, we use the standard orthonormal basis \cite{Benavides-Gallego:2018odl,Toshmatov:2015npp}
\begin{equation}\label{eq:A21}
\begin{aligned}
     e_{(t)}^{\mu} & =\frac{1}{\sqrt{\Delta \Sigma}}\left(a^{2}+r^{2}, 0,0, a\right), \\
e_{(r)}^{\mu} & =\frac{\sqrt{\Delta}}{\sqrt{\Sigma}}(0,1,0,0), \\
e_{(\theta)}^{\mu} & =\frac{1}{\sqrt{\Sigma}}(0,0,1,0), \\
e_{(\phi)}^{\mu} & =-\frac{1}{\sqrt{\Sigma \sin ^{2} \theta}}\left(a \sin ^{2} \theta, 0,0,1\right).
    \end{aligned}
\end{equation}
The components of the energy-momentum tensor derived from Eqs.(\ref{eq:A19}), (\ref{eq:A21}), and the Einstein field equation $ G_{\mu \nu} = 8 \pi T_{\mu \nu} $ are as follows:
\begin{equation}\label{eq:A22}
    \begin{aligned}
    T_{(t)(t)} & =\frac{1}{2} e_{(t)}^{\mu} e_{(t)}^{\nu} G_{\mu \nu}=\epsilon ,\hspace{0.3cm}
    T_{(\theta)(\theta)} =\frac{1}{2} e_{(\theta)}^{\mu} e_{(\theta)}^{\nu} G_{\mu \nu}=p_{\theta}, \\
 T_{(r)(r)}  &=\frac{1}{2} e_{(r)}^{\mu} e_{(r)}^{\nu} G_{\mu \nu}=p_{r}, \hspace{0.1cm}
T_{(\phi)(\phi)}  =\frac{1}{2} e_{(\phi)}^{\mu} e_{(\phi)}^{\nu} G_{\mu \nu}=p_{\phi},
    \end{aligned}
\end{equation}
from which we obtain
\begin{equation}\label{eq:A23}
\begin{aligned}
    \epsilon &= -p_{r}=T_{(t)(t)}=\frac{1}{4 \pi} \frac{r^{2} \rho^{\prime}}{\Sigma^{2}},
\end{aligned}
\end{equation}
\begin{equation}\label{eq:A24}
    \begin{aligned}
    p_{\theta} &= p_{\phi}=T_{(\theta)(\theta)}= \omega \left(\epsilon-\frac{2 \rho^{\prime}+r \rho^{\prime \prime}}{8 \pi \Sigma}\right).
    \end{aligned}
\end{equation}
When $Q=0$ and $K=\omega =0$, Eq.(\ref{eq:A24}) corresponds to rotating black holes with an anisotropic matter field and non-linear magnetic-charged black hole, respectively.
\subsection{Weak energy condition }
\begin{figure*}
	\begin{tabular}{c c}
		\includegraphics[scale=0.84]{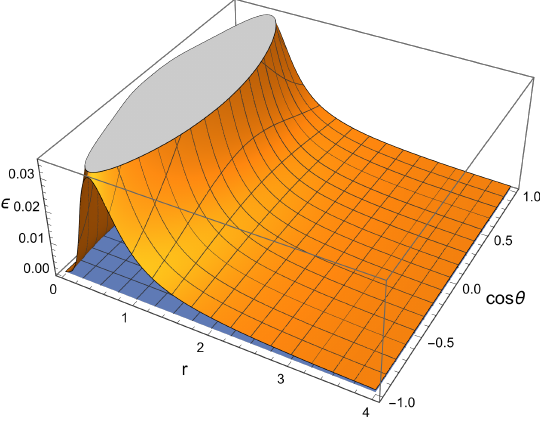}\hspace{-0.1cm}
		&\includegraphics[scale=0.92]{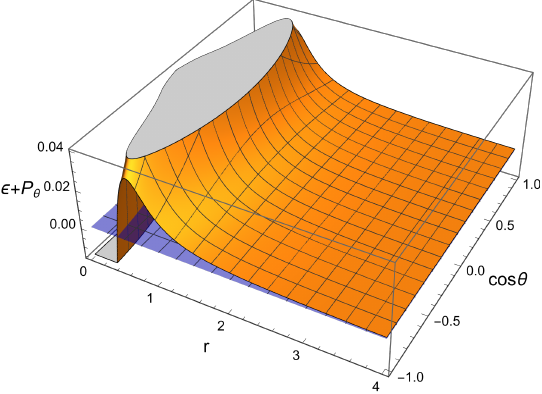}
	\end{tabular}
	\caption{Dependence of matter density $\epsilon$ and $\epsilon+p_{\theta}$ on radius and angle for a rotating and nonlinear magnetic-charged black hole with an anisotropic matter field for $M=1$, $Q=0.5$, $a=0.9$, $K=1$, $\omega =0.2$.}\label{fig:energy condition2}
\end{figure*}
The weak energy condition ensures that the behavior of matter and energy is physically reasonable, preventing phenomena such as negative energy density. It states that for any timelike vector $\xi^{\mu}$ at any point in spacetime, the stress-energy tensor $T_{\mu \nu}$ satisfies
\begin{equation}\label{eq:A25}
    \begin{aligned}
T_{\mu \nu} \xi^{\mu} \xi^{\nu} \geq 0.
    \end{aligned}
\end{equation}
With the decomposition of the energy-momentum tensor $T_{\mu \nu}$, the weak energy condition is equivalent to
\begin{equation}\label{eq:A26}
    \begin{aligned}
\epsilon \geq 0, \quad \epsilon+p_{i} \geq 0,
    \end{aligned}
\end{equation}
where $i=r$, $\theta$, $\varphi$. Substituting Eqs.(\ref{eq:A18}), (\ref{eq:A23}) and (\ref{eq:A24}) into Eq.(\ref{eq:A26}), we obtain
\begin{center}
\begin{equation}\label{eq:A27}
\begin{aligned}
    \epsilon &=\frac{r^2 \left(K (1-2 \omega ) r^{-2 \omega }+\frac{6 M Q^3 r^2}{\left(Q^3+r^3\right)^2}\right)}{8\pi\left(r^2+ a^2 \cos ^2\theta\right)^2} \geq 0,
    \end{aligned}
\end{equation}
\end{center}
\begin{center}
\begin{equation}\label{eq:A28}
\begin{aligned}
    \epsilon+p_{r}=0 ,
    \end{aligned}
\end{equation}
\end{center}
\begin{equation}\label{eq:A29}
    \begin{aligned}
    \epsilon+p_{\theta}&=\epsilon+p_{\phi}\\
     &= \omega \epsilon+\frac{3 M Q^3 r^2 \omega  \left(r^3-2 Q^3\right)}{4\pi\left(Q^3+r^3\right)^3 \left(a^2 \cos ^2\theta +r^2\right)}\\
    &-\frac{K \omega  \left(2 \omega ^2-3 \omega +1\right) r^{-2 \omega }}{8\pi(a^2 \cos ^2\theta +r^2)}\geq 0.
    \end{aligned}
\end{equation}
From Eqs.(\ref{eq:A27}), (\ref{eq:A28}) and (\ref{eq:A29}), we present the variations of  $\epsilon$ and $\epsilon+p_{\theta}/\epsilon+p_{\varphi}$ with $r$ and $\cos\theta$ under two sets of parameters in Figs.~\ref{fig:energy condition1} and \ref{fig:energy condition2}. It can be obtained from Figs.~\ref{fig:energy condition1} and \ref{fig:energy condition2} that the weak energy condition is violated of a rotating and non-linear magnetic-charged black hole with an anisotropic matter field.

Based on the analysis of parameter $\omega $ in Ref. \cite{Kim:2019hfp}, we know that when $0 \leq \omega  \leq \frac{1}{2}$, the energy density is not sufficiently localized, causing the total mass to diverge as $r \to \infty$. As a result, subsequent studies primarily focus on systems with $\omega  > \frac{1}{2}$ to ensure that the spacetime asymptotically approaches flatness at infinity.

\section{Geometric properties}
\label{sec:4}
\subsection{Horizons}
\begin{figure*}
		\centering
	\begin{tabular}{c c}
		\includegraphics[scale=0.55]{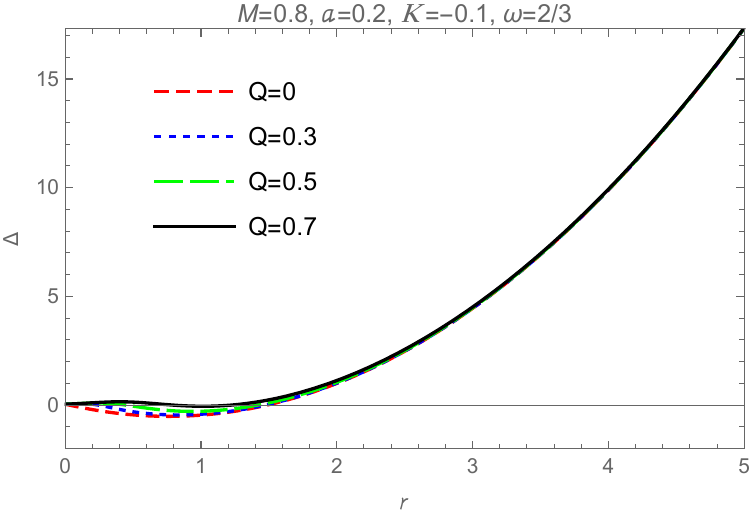}\hspace{-0.1cm}
		&\includegraphics[scale=0.54]{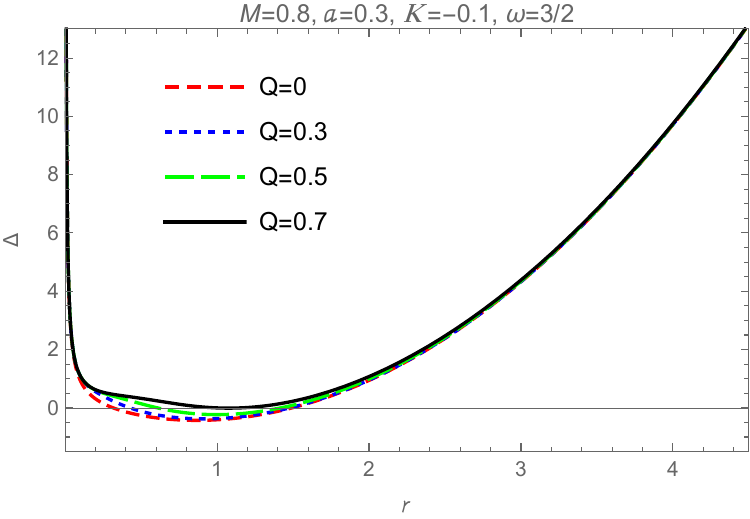}\\
        \includegraphics[scale=0.55]{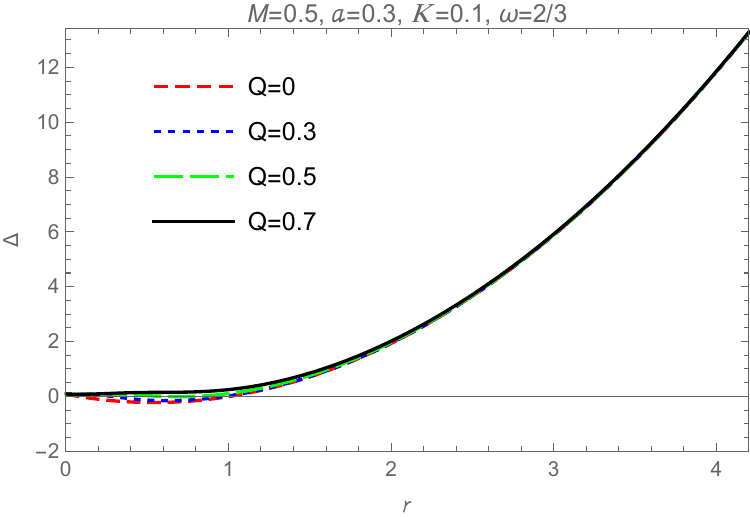}\hspace{-0.1cm}
		&\includegraphics[scale=0.55]{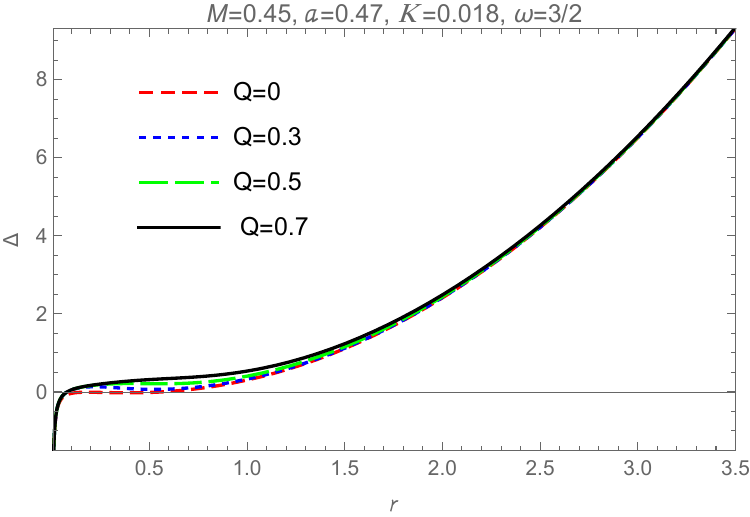}\\
	\end{tabular}
	\caption{The behavior of $\Delta$ vs $r$ for parameters $a$, $K$, and $\omega $ when $Q$ takes the values of 0, 0.3, 0.5, and 0.7.}\label{fig:event horizon1}
\end{figure*}

The event horizon corresponds to the Killing horizon, which defines the "surface" of a black hole as a three-dimensional hypersurface. The line element Eq.(\ref{eq:A17}) is singular at $\Delta= 0$, which corresponds to the event horizon of the black hole and satisfies
\begin{equation}\label{eq:B1}
    \begin{aligned}
      \Delta & =r^{2}+a^{2}-\frac{2 M r^{4}}{r^{3}+Q^{3}}-K r^{2\left(1-\omega \right)}=0 .
    \end{aligned}
\end{equation}
Based on the above equation, we show in Fig.~\ref{fig:event horizon1} the behavior of $\Delta$ vs $r$ for different parameters $a$, $K$, and $\omega $ with $Q$ taking values of $0$, $0.3$, $0.5$, and $0.7$, respectively. In Fig.~\ref{fig:event horizon1}, the first and second graphs correspond to the case of $K<0$, while the third and fourth graphs correspond to the case of $K>0$. From the first and third graphs in Fig.~\ref{fig:event horizon1}, for $\frac{1}{2} < \omega  < 1$, the influence of $Q$ on $\Delta$ is weakened as $r$ increases, and this influence becomes negligible when $r$ is large. This effect is noticeable within a small interval starting from $0$. From the second and fourth graphs in Fig.~\ref{fig:event horizon1}, for $\omega  > 1$, the influence of $Q$ on $\Delta$ is insignificant for both small and large values of $r$, with a noticeable effect only within a small interval of $r$. Therefore, the influence of $Q$ on $\Delta$ is mainly within a small interval. The above analysis shows that we can analyze the possible roots of $\Delta$ by considering the case when $Q = 0$ \cite{Kim:2019hfp}. By analyzing Eq.(\ref{eq:B1}) and Fig.~\ref{fig:event horizon1}, we can find the following conclusions:
\begin{itemize} 	
	\item{For $\frac{1}{2} < \omega  < 1$, $\Delta$ is mainly dominated by the $r^2$ term for large $r$, while for smaller $r$, it is mainly influenced by $r^2$ and the non-linear magnetic-charged term. The influence of the nonlinear magnetic charge term causes $\Delta$ ($Q \neq 0$) to deviate from the case $\Delta$ ($Q = 0$) in a small interval starting from $r = 0$. For both cases $K < 0$ and $K > 0$, the rotating black hole may have two horizons: the Cauchy horizon and the event horizon. For example, the first and third graphs in Fig.~\ref{fig:event horizon1}, $\Delta$ graphs correspond to the cases of $K = -0.1$, $\omega  = \frac{2}{3}$ and $K = 0.1$, $\omega  = \frac{2}{3}$, respectively.
}
	\item {For $\omega>1$, $\Delta$ is mainly dominated by the $r^2$ term for large $r$, while for smaller $r$, it is primarily influenced by the anisotropic matter field term. The influence of the non-linear magnetic-charged term causes $\Delta$ ($Q \neq 0$) to deviate within a small range of $r$ compared to the case $\Delta$ ($Q = 0$).
For $K < 0$, $\Delta$ starts from positive infinity at $r \to 0$, decreases rapidly, and then increases again. Therefore, for $\omega  >1$ and $K < 0$, the black hole may have two horizons. For $K > 0$, $\Delta$ starts from negative infinity at $r \to 0$, increases rapidly, then decreases and then increases again, or $\Delta$ starts from negative infinity at $r \to 0$, increases rapidly and then increases with the $r^2$ term. Therefore, for $\omega  >1$ and $K > 0$, the black hole may have three horizons. For example, the second and fourth graphs in Fig.~\ref{fig:event horizon1}, they correspond to $\Delta$ images for $K = -0.1$, $\omega  = \frac{3}{2}$ and $K = 0.018$, $\omega  = \frac{3}{2}$, respectively.}
\end{itemize}
\subsection{Ergosphere}
\begin{figure*}
		\centering
	\begin{center}
		\begin{tabular}{c c c c}
			\includegraphics[scale=0.48]{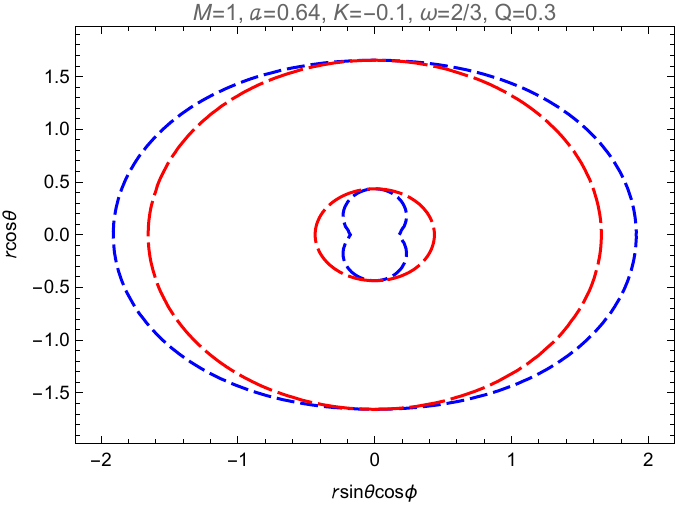}\hspace{0cm}
			\includegraphics[scale=0.48]{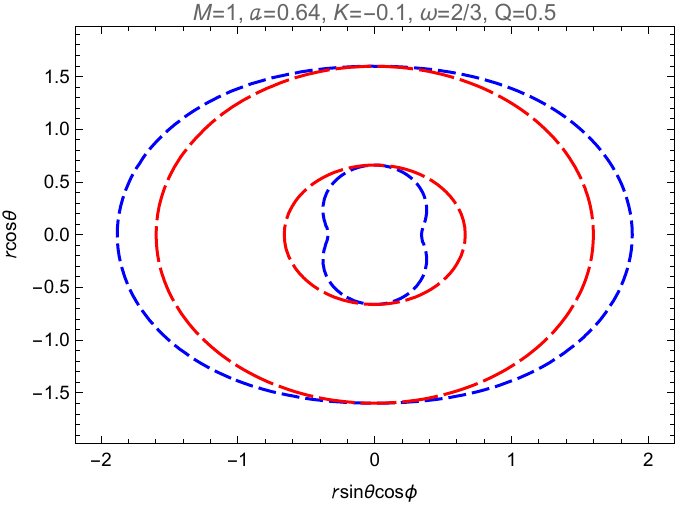}\hspace{0cm}
			&\includegraphics[scale=0.48]{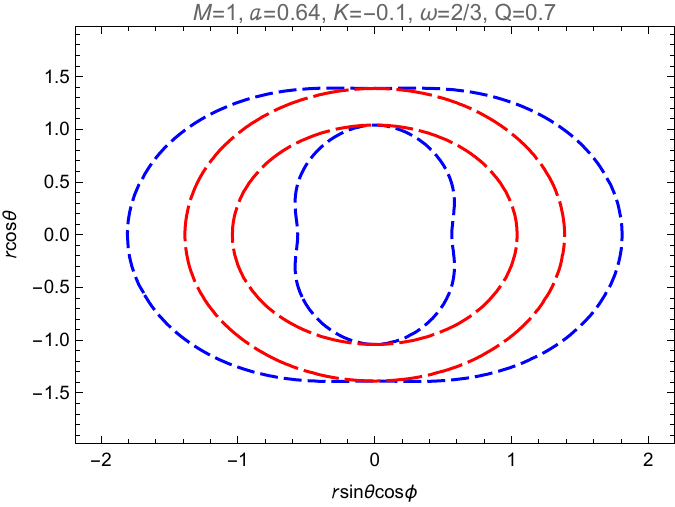}\\
			\includegraphics[scale=0.48]{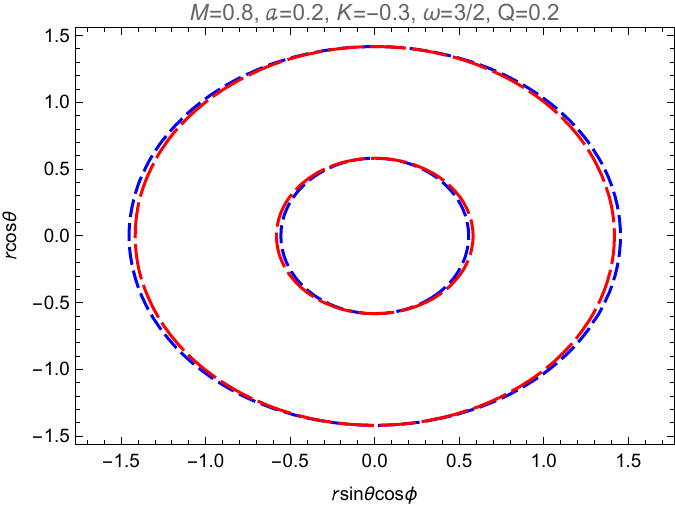}\hspace{0cm}
			\includegraphics[scale=0.48]{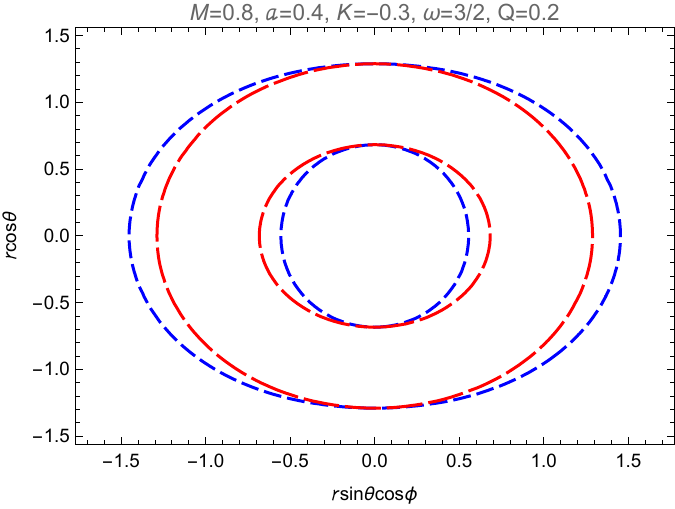}\hspace{0cm}
			&\includegraphics[scale=0.48]{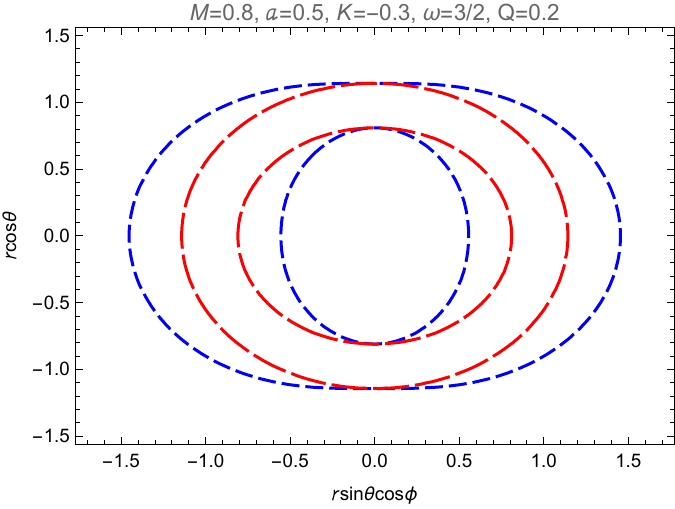}\\
			\includegraphics[scale=0.48]{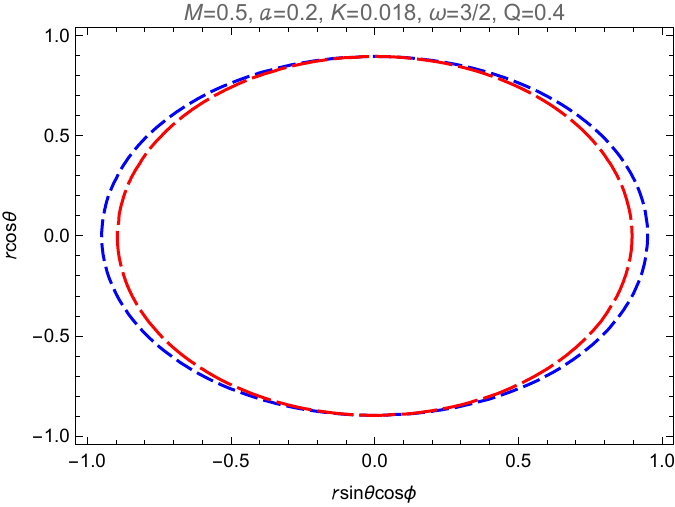}\hspace{0cm}
			\includegraphics[scale=0.48]{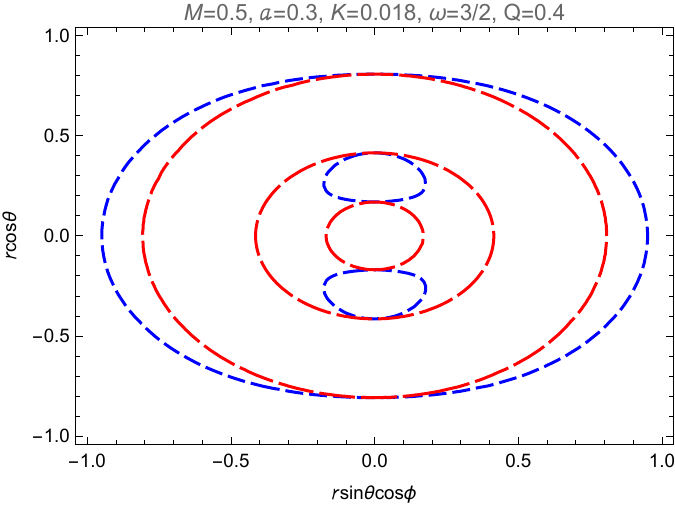}\hspace{0cm}
			&\includegraphics[scale=0.48]{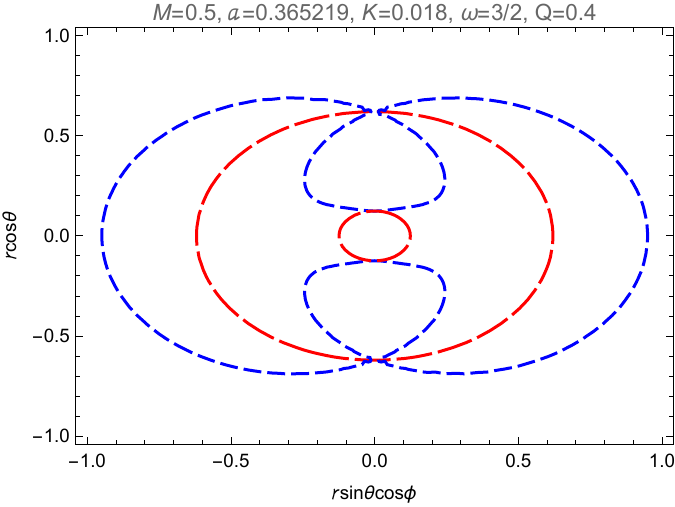}\\
		\end{tabular}
		\caption{The behavior of ergosphere in the xz-plane for different parameters $M$, $a$, $K$, $\omega $ and $Q$.}
		\label{fig:ergosphere}
	\end{center}
\end{figure*}

The stationary limit surface, also known as the infinite redshift surface, is where the timelike Killing vector $K^\mu = \partial_t$ satisfies $K^\mu K_\mu = 0$. Therefore, we obtain
\begin{equation}\label{eq:B2}
    \begin{aligned}
     r^{2}+a^{2}\cos ^2\theta-\frac{2 M r^{4}}{r^{3}+Q^{3}}-K r^{2\left(1-\omega \right)}=0 .
    \end{aligned}
\end{equation}
The ergosphere is a region that lies outside the black hole, bounded by the event horizon and the outer stationary limit surface. The extraction of energy from a rotating black hole relies on the ergosphere region, which corresponds to the area between the outermost red curve and the blue curve in each graph in Fig.~\ref{fig:ergosphere}. Using Eqs.(\ref{eq:B1}) and (\ref{eq:B2}), we present the behavior of the ergosphere in the xz-plane for different parameters $a$, $K$, $\omega $ and $Q$ in Fig.~\ref{fig:ergosphere}. In the first row of ergosphere graphs in Fig.~\ref{fig:ergosphere}, we show that when other parameters are kept constant and parameter $Q$ is changed, the area of the ergosphere increases with the increase of parameter $Q$. In the second and third rows of ergosphere graphs in Fig.~\ref{fig:ergosphere}, we show that when other parameters are kept constant and parameter $a$ is changed, the area of the ergosphere increases with the increase of parameter $a$. In the third row of ergosphere graphs in Fig.~\ref{fig:ergosphere}, we show that for $K > 0$ and $\omega  >1$, as parameter $a$ increases, the black hole transitions from having a single horizon to three horizons, and then to two horizons.

\section{Thermodynamics of the rotating black hole}
\label{sec:5}

We now know that a black hole is not just a gravitational system, but also a thermodynamic system. This understanding originates from the proposal of black hole entropy \cite{Bekenstein:1973ur,Bekenstein:1974jk} and Hawking temperature \cite{Hawking:1974rv,Hawking:1975vcx}. The Bekenstein-Hawking entropy is one-quarter of the event horizon area. Therefore, the entropy of a rotating and  non-linear magnetic-charged black hole with an anisotropic matter field is given by
\begin{equation}\label{eq:C1}
    \begin{aligned}
     S=\frac{\mathcal{A}}{4}=\pi\left(r_{H}^{2}+a^{2}\right),
    \end{aligned}
\end{equation}
where $r_{H}$ is the event horizon, satisfying Eqs.(\ref{eq:B1}). On the event horizon, the Hawking temperature is proportional to the surface gravity $\kappa$:
\begin{equation}\label{eq:C2}
    \begin{aligned}
    T&=\frac{\kappa}{2 \pi}=\frac{\partial_{r_{H}} \Delta\left(r_{H}\right)}{4 \pi\left(r_{H}^{2}+a^{2}\right)}\\
     &=\frac{1}{4 \pi  r_{H} \left(a^2+r_{H}^2\right) \left(Q^3+r_{H}^3\right)}(-a^2 \left(4 Q^3+r_{H}^3\right)+r_{H}^5\\
     &+K r_{H}^{-2 \omega } \left(2 Q^3 r_{H}^2 (\omega +1)+r_{H}^5 (2 \omega -1)\right)-2 Q^3 r_{H}^2).
    \end{aligned}
\end{equation}
The angular velocity of a rotating black hole at the event horizon is
\begin{equation}\label{eq:C3}
    \begin{aligned}
    \Omega=-\left.\frac{g_{t \phi}}{g_{\phi \phi}}\right|_{r=r_{H}}=\frac{a}{r_{H}^{2}+a^{2}} .
    \end{aligned}
\end{equation}
In the above, we directly derived the entropy, temperature, and angular velocity of the rotating black hole through the relationship between thermodynamic quantities and classical mechanical quantities. Nevertheless, for other thermodynamic quantities of the black hole, such as the electric potential, it is not easy to solve using the above similar methods. Below, we use the first law of thermodynamics and a squared-mass formula of the rotating and non-linear magnetic-charged black hole with an anisotropic matter field to solve for the various thermodynamic quantities of the rotating black hole.

For the rotating and non-linear magnetic-charged black hole with an anisotropic matter field, we take $S$ and $Q$, $J$, $r_{0}$ as four independent thermodynamic quantities and present its first law of thermodynamics as \cite{Smarr:1972kt}
\begin{equation}\label{eq:C4}
    \begin{aligned}
    d M=T d S+\Phi d Q+\Omega d J+\Phi_{0} d r_{0} ,
    \end{aligned}
\end{equation}
where $\Phi$ is the magnetic charge potential for $Q$, $J$ is the angular momentum and $\Phi_{0}$ is the electric potential for $r_0$.
From Eqs.(\ref{eq:B1}), (\ref{eq:C2}) and $J=Ma$, we obtain the squared-mass formula for a rotating and non-linear magnetic-charged black hole with an anisotropic matter field as \cite{Christodoulou:1970wf,Christodoulou:1971pcn}
\begin{equation}\label{eq:C5}
    \begin{aligned}
    M^2&=\frac{\pi }{S}J^2+\frac{\pi}{4S}\left(\frac{(\frac{S}{\pi}-a)^{\frac{3}{2}}+Q^{3}}{(\frac{S}{\pi}-a^{2})^{\frac{3}{2}}}\right)^{2}\\
     &\times \left(\frac{S}{\pi}
    -\frac{r_{0}^{2\omega }}{(1-2\omega )}\left(\frac{S}{\pi}-a^{2}\right)^{\frac{1}{2}\left(1+\frac{r_{0}^{2\omega }}{K}\right)}\right)^{2}.
    \end{aligned}
\end{equation}
From Eqs. (\ref{eq:C4}) and (\ref{eq:C5}), we obtain the following two thermodynamic quantities:
\begin{equation}\label{eq:C6}
    \begin{aligned}
    \Phi&=\left(\frac{\partial M}{\partial Q}\right)_{S,J,r_{0}}\\
     &=\frac{3Q^2}{2r_{H}^2(r_{H}^2+a^2)}\left(r_{H}^2+a^2-Kr_{H}^{2(1-\omega )}\right),
    \end{aligned}
\end{equation}
and
\begin{equation}\label{eq:C7}
    \begin{aligned}
    \Phi_{0}&=\left(\frac{\partial M}{\partial r_{0}}\right)_{S,Q,J}\\
     &=\frac{\omega }{2\omega -1}\left(\frac{r_{H}^{3}+Q^{3}}{r_{H}^{3}}\right)\left(\frac{r_{H}r_{0}}{r_{H}^{2}+a^{2}}\right)\left(\frac{r_{0}}{r_{H}}\right)^{2(\omega -1)}.
    \end{aligned}
\end{equation}
For Eq.(\ref{eq:C7}), when $Q=a=0$ and $Q =0$ ($a\ne 0$), it degenerates into a Schwarzschild black hole with an anisotropic matter field \cite{Cho:2017nhx} and rotating black hole with an anisotropic matter field \cite{Kim:2019hfp}, respectively.
For a thermodynamic system of a black hole, the thermodynamic stability of the black hole can be studied by its heat capacity. Therefore, we obtain the heat capacity of a rotating and non-linear magnetic-charged black hole with an anisotropic matter field and analyze the stability of the rotating black hole. The heat capacity $C$ calculated at constant angular momentum and charge in the canonical ensemble can be obtained as follows:
\begin{align}\label{eq:C8}
C &= T\frac{\partial S}{\partial T} \nonumber \\
  &= 2 \pi  \left(a^2+r_{H}^2\right) \left(Q^3+r_{H}^3\right) r_{H}^{2 \omega +2} [-a^2 \left(4 Q^3+r_{H}^3\right) \nonumber \\
  &+K r_{H}^{-2 \omega } \left(2 Q^3 r_{H}^2 (\omega +1)+r_{H}^5 (2 \omega -1)\right)-2 Q^3 r_{H}^2+r_{H}^5] \nonumber \\
  &/[a^4 \left(4 Q^6+14 Q^3 r_{H}^3+r_{H}^6\right) r_{H}^{2 \omega } \nonumber \\
  &+a^2 r_{H}^2 [2 \left(5 Q^6+16 Q^3 r_{H}^3+2 r_{H}^6\right) r_{H}^{2 \omega } \nonumber \\
  &-K [2 Q^6 \left(2 \omega ^2+\omega -1\right)+2 Q^3 r_{H}^3 \left(4 \omega ^2-\omega +4\right) \nonumber \\
  &+r_{H}^6 (1-2 \omega )^2]]-r_{H}^4 [K [Q^6 \left(4 \omega ^2+6 \omega +2\right) \nonumber \\
  &+2 Q^3 r_{H}^3 \left(4 \omega ^2+3 \omega +5\right)+r_{H}^6 \left(4 \omega ^2-1\right)] \nonumber \\
  &+\left(-2 Q^6-10 Q^3 r_{H}^3+r_{H}^6\right) r_{H}^{2 \omega }]].
\end{align}
\begin{figure*}
		\centering
	\begin{tabular}{c c}
		\includegraphics[scale=0.55]{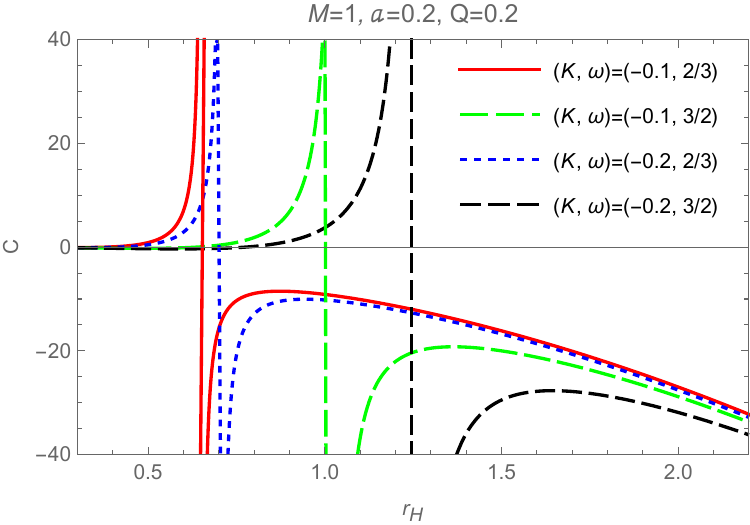}\hspace{-0.1cm}
		&\includegraphics[scale=0.55]{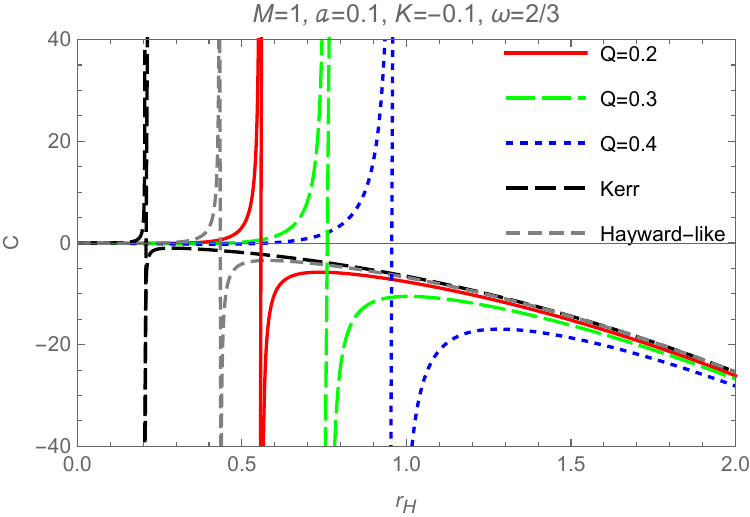}\\
	\end{tabular}
	\caption{The behavior of the heat capacity vs $r_{H}$ for different parameters $a$, $K$, $\omega $ and $Q$.}\label{fig:Heat capacity}
\end{figure*}
When $Q = 0 $, the above equation degenerates into the heat capacity of a rotating black hole with an anisotropic matter field \cite{Kim:2019hfp}. Based on Eq.(\ref{eq:C8}), we plot the behavior of heat capacity vs $r_{H}$ for different parameters $Q$, $a$, $K$ and $w$ in Fig.~\ref{fig:Heat capacity}. For a black hole thermodynamic system, the positive and negative values of the black hole heat capacity correspond to local thermodynamic stability and instability, respectively. In Fig.~\ref{fig:Heat capacity}, the left and right graphs correspond to the heat capacity graphs under different ($K$, $\omega $) parameters and different $Q$ parameter, respectively. We can observe that when the black hole event horizon radius satisfies $0 < r_{H} < r_{H}^{C}$, the system is thermodynamically stable, while for $r_{H} > r_{H}^{C}$, it is thermodynamically unstable. At $r_{H}=r_{H}^{C}$, the black hole heat capacity diverges, which indicates a second-order phase transition.
\section{Penrose process}
\label{sec:6}
In 1969, British physicist Penrose proposed a theoretical process for extracting energy from a rotating black hole, known as the Penrose process. Here, we calculate the Penrose process for a rotating and non-linear magnetic-charged black hole with an anisotropic matter field \cite{Haroon:2017opl,Bhattacharya:2017scw,Vertogradov:2022eeq}, and obtain the energy gain and extraction efficiency of this process.

The basic idea of the Penrose process is that a particle $A$ moving on the equatorial plane ($\theta =\pi/2$, $u^\theta =0$) of the black hole entering the ergosphere, where it decays into two massless particles, $B$ and $C$. The massless particle $B$ falls into a negative energy orbit, while massless particle $C$ escapes to infinity. According to the energy conservation, the energy of massless particle $C$ exceed the initial energy of particle $A$.
The four-momentum of a particle is
\begin{equation}\label{eq:PT1}
	\begin{array}{c}
		\begin{aligned}
			P_{\mu} &=g_{\mu \nu} u^{\nu} ,
		\end{aligned}
	\end{array}
\end{equation}
where $u^{\nu}$ is the four-velocity of a particle.
Considering that particles moving on the equatorial plane of a black hole satisfy energy and momentum conservation, we obtain
\begin{equation}\label{eq:D1}
	\begin{array}{c}
		\begin{aligned}
			P_{t} &=-E=g_{t t} u^{t}+g_{t \phi} u^{\phi  } ,
		\end{aligned}
	\end{array}
\end{equation}
\begin{equation}\label{eq:D2}
	\begin{array}{c}
		\begin{aligned}
			P_{\phi} &=L=g_{\phi \phi} u^{\phi  }+g_{t \phi} u^{t},
		\end{aligned}
	\end{array}
\end{equation}
where $E$ and $L$ are the energy and angular momentum of the test particle, respectively. The radial equation for the geodesic motion of a test particle, derived from Eqs. (\ref{eq:D1}) and (\ref{eq:D2}) and using the condition $P_{\nu} P^{\nu}=-\delta$ where $\delta=-1$, $0$, $1$ correspond to time-like, null, and space-like geodesics, respectively, is as follows:
\begin{equation}\label{eq:D3}
	\begin{array}{c}
		\begin{aligned}
			r^{4} u^{r}&=E^{2}\left(r^{4}+a^{4}+a^{2}\left(2 r^{2}-\Delta\right)\right)\\
     &+L^{2}\left(a^{2}-\Delta\right)-4 a L E \rho r+\delta r^{2} \Delta.
		\end{aligned}
	\end{array}
\end{equation}
\begin{figure*}
		\centering
	\begin{tabular}{c c}
		\includegraphics[scale=0.55]{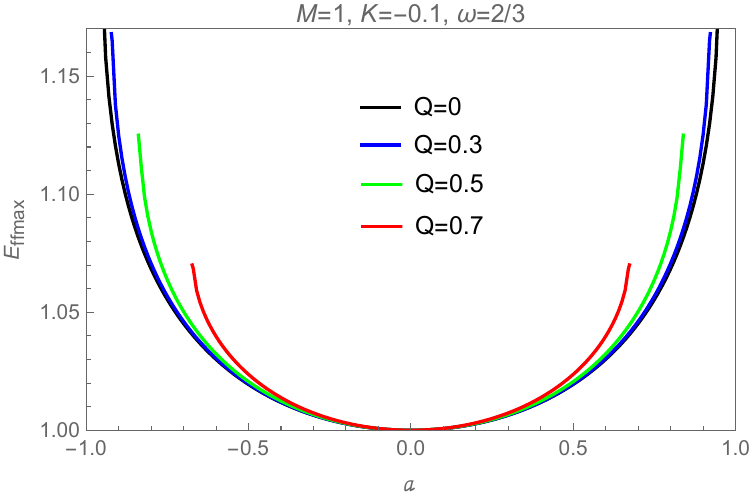}\hspace{-0.1cm}
		&\includegraphics[scale=0.55]{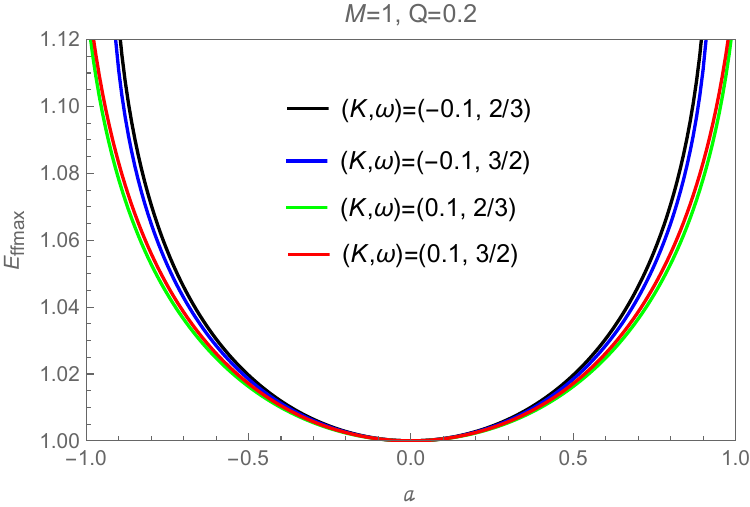}\\
	\end{tabular}
	\caption{The behavior of maximal efficiency of Penrose process vs $a$ for different parameters $K$, $\omega $ and $Q$.}\label{fig:maximal efficiency}
\end{figure*}
Assuming that the decay of particle $A$ occurs at the turning point of the geodesic ($u^{r}=0$), so we can derive from Eq. (\ref{eq:D3}) and the condition $u^{r} = 0$ that
\begin{equation}\label{eq:D4}
	\begin{array}{c}
		\begin{aligned}
			E=\frac{2 a \rho r L \pm r \sqrt{\Delta} \sqrt{L^{2} r^{2}-\delta\left(r^{4}+a^{4}+a^{2}\left(2 r^{2}-\Delta\right)\right)}}{r^{4}+a^{4}+a^{2}\left(2 r^{2}-\Delta\right)},
		\end{aligned}
	\end{array}
\end{equation}
and
\begin{equation}\label{eq:D5}
	\begin{array}{c}
		\begin{aligned}
			L=\frac{-2 a \rho r E \pm r \sqrt{\Delta} \sqrt{E^{2} r^{2}+\delta\left(r^{2}-2 \rho r\right)}}{\Delta-a^{2}} .
		\end{aligned}
	\end{array}
\end{equation}
In order to have negative energy, the angular momentum must satisfy $L < 0$. Therefore, from Eq. (\ref{eq:D5}), we obtain
\begin{equation}\label{eq:D6}
	\begin{array}{c}
		\begin{aligned}
			1-\frac{2 \rho}{r}<\delta \frac{\Delta}{L^{2}} .
		\end{aligned}
	\end{array}
\end{equation}
Given that $g_{tt}(\theta=\frac{\pi}{2})=-\left(1 - \frac{2\rho}{r}\right) $, it is evident from the inequality that this phenomenon can occur within the ergosphere. We consider a particle $A$ ($\delta=-1$) with energy $E_A=1>0$ and angular momentum $L=L_A$ entering the ergosphere. It then decays into two massless particles $B$ and $C$ ($\delta=0$), with energies and angular momenta ($E_B$, $L_B$) and ($E_C$, $L_C$), respectively. The massless particle $B$ that falls into the event horizon has negative energy, while the massless particle $C$ that escapes to infinity has positive energy. From Eq.(\ref{eq:D5}), we obtain the relationship between the angular momentum and energy for $A$, $B$, and $C$ as
\begin{equation}\label{eq:D7}
	\begin{array}{c}
		\begin{aligned}
			L_{A} & =\frac{-2 a \rho r+r \sqrt{\Delta} \sqrt{2 \rho r}}{\Delta-a^{2}}, \\
L_{B} & =\frac{-2 a \rho r-r^{2} \sqrt{\Delta}}{\Delta-a^{2}} E_{B}, \\
L_{C} & =\frac{-2 a \rho r+r^{2} \sqrt{\Delta}}{\Delta-a^{2}} E_{C}.
		\end{aligned}
	\end{array}
\end{equation}
The conservation of energy and angular momentum is given as follows:
\begin{equation}\label{eq:D8}
	\begin{array}{c}
		\begin{aligned}
			E_{B}+E_{C}=E_{A}=1,
		\end{aligned}
	\end{array}
\end{equation}
and
\begin{equation}\label{eq:D9}
	\begin{array}{c}
		\begin{aligned}
			L_{B}+L_{C}=L_{A}.
		\end{aligned}
	\end{array}
\end{equation}
The energies of two massless particles $B$ and $C$ obtained from Eqs. (\ref{eq:D7}), (\ref{eq:D8}) and (\ref{eq:D9}) are given by
\begin{equation}\label{eq:D10}
	\begin{array}{c}
		\begin{aligned}
			E_{B}&=\frac{1}{2}\left(1-\sqrt{1+\frac{a^{2}-\Delta}{r^{2}}}\right)\\
                =&\frac{1}{2}\left(1-\sqrt{\frac{2 M r^{2}}{r^{3}+Q^{3}}+K r^{-2\omega }}\right),
		\end{aligned}
	\end{array}
\end{equation}
and
\begin{equation}\label{eq:D11}
	\begin{array}{c}
		\begin{aligned}
              E_{C}=&\frac{1}{2}\left(1+\sqrt{1+\frac{a^{2}-\Delta}{r^{2}}}\right)\\
              =&\frac{1}{2}\left(1+\sqrt{\frac{2 M r^{2}}{r^{3}+Q^{3}}+K r^{-2\omega }}\right) .
		\end{aligned}
	\end{array}
\end{equation}
From the above equation, we find that the energy of the massless particle $C$ escaping to infinity is higher than the energy of the particle $ A$ that initially entered the ergosphere of the black hole. The energy gain $\Delta E$ can be obtained as
\begin{equation}\label{eq:D12}
	\begin{array}{c}
		\begin{aligned}
              \Delta E=\frac{1}{2}\left(\sqrt{\frac{2 M r^{2}}{r^{3}+Q^{3}}+K r^{-2\omega }}-1\right)=-E_{B} .
		\end{aligned}
	\end{array}
\end{equation}
According to Eq.(\ref{eq:D10}), the maximum energy gain occurs at the event horizon, and the maximal efficiency of the Penrose process is then given by
\begin{equation}\label{eq:D13}
	\begin{array}{c}
		\begin{aligned}
              E_{ff \max }&=\frac{E_{A}+\Delta E}{E_{A}}\\
              &=\frac{1}{2}\left(1+\sqrt{\frac{2 M r_{H}^{2}}{r_{H}^{3}+Q^{3}}+K r_{H}^{-2\omega }}\right).
		\end{aligned}
	\end{array}
\end{equation}
 The behavior of the maximal efficiency of the Penrose process vs $a$ for different parameters $K$, $\omega$ and $Q$ is shown in Fig.~\ref{fig:maximal efficiency}. We observe that $E_{f f \max }$ increases with the increase of the spin parameter $a$ and the magnetic charge $Q$. The influence of the anisotropic matter field on the $E_{f f \max }$ is realized through its effect on the event horizon $r_{H}$.		

\section{CONCLUSION AND DISCUSSION }
\label{sec:7}
In this article, we have first presented the exact solution of the static spherically symmetric non-linear magnetic-charged black hole with an anisotropic matter field and further calculated the curvature of the black hole. The results show that the curvature diverges at $r=0$, indicating that the non-linear magnetic-charged black hole with an anisotropic matter field has a singularity. We have used a modified Newman-Janis algorithm to obtain the solution for a rotating and non-linear magnetic-charged black hole with an anisotropic matter field. Subsequently, we have studied the energy-momentum tensor and the weak energy condition for the rotating black hole, finding that the rotating black hole violates the weak energy condition.

We have investigated the geometric properties and thermodynamic properties of a rotating and nonlinear magnetic-charged black hole with an anisotropic matter field. Regarding geometric properties, we first have analyzed the influence of the spin parameter $a$ and the magnetic charge $Q$ on $\Delta$ in Fig.~\ref{fig:event horizon1} and have found that the influence of $Q$ is mainly within a small interval. Further analysis of Fig.~\ref{fig:event horizon1} and $\Delta$ ($Q = 0$) \cite{Kim:2019hfp} have revealed that the rotating black hole may have two horizons for $\frac{1}{2} < \omega < 1$, regardless of whether $K$ is positive or negative. For $\omega  > 1$ and $K < 0 $, $\Delta$ can have at most two roots, while for $\omega  > 1$ and $K > 0 $, the black hole can have at most three roots, indicating the possible existence of three horizons. Further studies on the influence of various parameters on the ergosphere of the black hole have showed that the area of the ergosphere increases with the spin parameter $a$ and the magnetic charge $Q$.

Regarding thermodynamic properties, we have obtained the entropy, Hawking temperature and angular velocity of the rotating and non-linear magnetic-charged black hole with an anisotropic matter field. To further derive other thermodynamic quantities, we have presented the first law of thermodynamics for the black hole. Using $\Delta= 0$, we have derived the squared mass formula and have calculated other thermodynamic quantities. We further have studied the stability of this black hole through its heat capacity and have found that the black hole is thermodynamically stable when the event horizon radius satisfies $0 < r_{H} < r_{H}^{C}$. However, when $r_{H} > r_{H}^{C}$, the black hole becomes thermodynamically unstable. At $r_{H} = r_{H}^{C}$, the heat capacity diverges, which indicates a second-order phase transition.

Finally, we have calculated the energy extraction efficiency of the Penrose process of a rotating and non-linear magnetic-charged black hole with an anisotropic matter field and have found that the maximal efficiency increases with the spin parameter $a$ and the magnetic charge $Q$. The influence of the anisotropic matter field on the maximum efficiency of the Penrose process is realized through its impact on the event horizon.

In this paper, we mainly analyze black hole properties under various parameter values without imposing constraints on these parameters. We know that the shape of the black hole shadow is sensitive to parameter changes. Combined with observational data from Event Horizon Telescope (EHT) and others, we can effectively eliminate parameter regions that do not match the measured results. In our subsequent research, we will constrain the parameter space of this black hole through the observational features of the black hole shadow, such as the axis ratio and angular diameter.

\end{document}